\documentclass[aps,prl,final,twocolumn,letterpaper]{revtex4}

\usepackage{graphicx}   
\usepackage{import}                         
\usepackage{epstopdf}
\usepackage{amsmath} 
\usepackage{float} 
\usepackage{bm}
\usepackage{amssymb}
\usepackage{quotes}
\usepackage{indentfirst}
\usepackage{color}
\usepackage[utf8]{inputenc}
\usepackage{transparent}
\usepackage{dcolumn}
\usepackage{braket}
\usepackage{multirow}
\usepackage{cancel} 
\usepackage{mdframed}
\usepackage{color}
\usepackage{bm}
\usepackage{dsfont}
\newcommand{\nv}{NV$^-$}
\newcommand{\nz}{NV$^0$}
\newcommand{\rabi}{\Omega_\mathrm{R}}
\newcommand{\gtwos}{\gamma_2^{\star}}
\newcommand{\gtwo}{\gamma_2}
\newcommand{\gone}{\gamma_1}

\usepackage{slashed}
\usepackage{graphicx}
\usepackage{amsmath}
\usepackage{siunitx}
\usepackage{hyperref}
\usepackage{soul, color} 
\soulregister\ref{7}  
\soulregister\cite{7} 
\renewcommand{\st}[1]{}

\usepackage{xr}

\makeatletter
\newcommand*{\addFileDependency}[1]{
  \typeout{(#1)}
  \@addtofilelist{#1}
  \IfFileExists{#1}{}{\typeout{No file #1.}}
}
\makeatother

\usepackage{textcomp} 
\usepackage{xifthen}
\usepackage{xcolor}
\usepackage{etoolbox}
\newboolean{togglechanges} 

\setboolean{togglechanges}{false}

\newcommand{\comment}[1]{\ifbool{togglechanges}
    {#1}  
    {\textcolor{blue}{#1}}}

\usepackage{bibentry}

\usepackage{graphicx}
\usepackage{dcolumn}
\usepackage{bm}

\usepackage{textcomp} 

\makeatletter
\renewcommand{\fnum@figure}{\textbf{Fig.~\thefigure}}
\makeatother

\begin{document}
\rmfamily

\title{Quantum sensing of electron beams using solid-state spins}


\author{Jakob M. Grzesik$^{\dagger,1}$}
\author{Dominic Catanzaro$^{\dagger,1}$}
\author{Charles Roques-Carmes$^{\dagger,1}$}%
\email{chrc@stanford.edu}   
\author{Eric~I.~Rosenthal$^{1}$}
\author{Guido~L.~van~de~Stolpe$^{1}$}
\author{Aviv~Karnieli$^{1}$}
\author{Giovanni~Scuri$^{1}$}
\author{Souvik~Biswas$^{1}$}
\author{Kenneth~J.~Leedle$^{1}$}
\author{Dylan~S.~Black$^{1}$}
\author{Robert~L.~Byer$^{1}$}
\author{Ido~Kaminer$^{2}$}
\author{R.~Joel~England$^{3}$}
\author{Shanhui~Fan$^{1}$}
\author{Olav~Solgaard$^{1}$}
\author{Jelena~Vu\v{c}kovi\'{c}$^{1}$}

\affiliation{\vspace{0.1cm}$^1$~E. L. Ginzton Laboratory, Stanford University, 348 Via Pueblo, Stanford, CA, USA\looseness=-1
\\
$^{2}$~Department of Electrical and Computer Engineering, Technion-Israel Institute of Technology, Haifa, Israel\looseness=-1 \\ 
$^{3}$~SLAC National Accelerator Laboratory, 2575 Sand Hill Road, Menlo Park, CA, USA\looseness=-1 \\ 
$^\dagger$ These authors contributed equally to the work.}



\clearpage 

\vspace*{-2em}



\begin{abstract}
Scattering experiments with energetic particles, such as free electrons, have been historically used to reveal the quantum structure of matter. However, realizing \textit{coherent} interactions between free-electron beams and solid-state quantum systems has remained out of reach, owing to their intrinsically weak coupling. Realizing such coherent control would open up opportunities for hybrid quantum platforms combining free electrons and solid-state qubits for coincident quantum information processing and nanoscale sensing. Here, we present a framework that employs negatively charged nitrogen-vacancy centers (\nv) in diamond as quantum sensors of a bunched electron beam. We develop a Lindblad master equation description of the magnetic free-electron--qubit interactions and identify spin relaxometry as a sensitive probe of the interaction. Experimentally, we integrate a confocal fluorescence microscopy setup into a microwave-bunched electron beam line. We monitor charge-state dynamics and assess their impact on key sensing performance metrics (such as spin readout contrast), defining safe operating parameters for quantum sensing experiments. By performing $T_1$ relaxometry under controlled electron beam exposure, we establish an upper bound on the free-electron--spin coupling strength. Our results establish \nv centers as quantitative probes of free electrons, providing a metrological benchmark for free-electron--qubit coupling under realistic conditions, and chart a route toward solid-state quantum control with electron beams.\end{abstract}

\maketitle

\section{Introduction}

Scattering experiments using high-energy particles have historically played a pivotal role in shaping our understanding of quantum mechanics by revealing the fundamental structure of matter~\cite{rutherford1911lxxix, geiger1910scattering}. A prime example is the Franck-Hertz experiment~\cite{franck1967zusammenstosse}, which more than a century ago demonstrated the quantization of atomic energy levels through electron scattering. Since then, remarkable progress has been made in engineering quantum systems, particularly through control of light-matter interactions -- a cornerstone of contemporary quantum technologies~\cite{reiserer2015cavity, degen2017quantum, o2009photonic, raimond2001manipulating, leibfried2003quantum}. Concurrently, electron microscopy has undergone tremendous advancements, now routinely offering ultrafast temporal resolutions, sub-nanometer spatial precision, and energy-loss spectroscopy at the millielectronvolt scale, enabling unprecedented characterization of atomic-scale phenomena~\cite{chen2021electron}, nanoscale electromagnetic effects~\cite{garcia2010optical, de2025roadmap, roques2023free, polman2019electron}, and ultrafast imaging of spatiotemporal electronic and polaritonic dynamics~\cite{hassan2017high, Kurman2021SpatiotemporalElectrons}. 

A promising yet underexplored avenue is to revisit such scattering experiments, not merely for fundamental measurements, but as a means to coherently probe and manipulate quantum systems. One recent proposal in this direction is the so-called ``free-electron--bound-electron resonant interaction'' (FEBERI~\cite{gover2020free}), a coherent analogue of the Franck-Hertz experiment. Several proposals suggest resonantly-modulated free electrons can enable nanoscale quantum control~\cite{gover2020free, garcia2022complete}, such as coherent manipulation of spin qubits~\cite{ratzel2021controlling}, novel microscopy modalities capable of imaging quantum coherence at previously unattainable resolutions~\cite{ruimy2021toward, karnieli2023quantum}, and the realization of multiqubit gates mediated by electron beams~\cite{zhao2021quantum}.

Despite these compelling theoretical proposals, the experimental observation of coherent interactions between modulated electron beams and quantum systems has remained elusive. Three primary roadblocks have impeded progress: first, the interaction strength is inherently weak, demanding large electron beam currents, near-perfect bunching at the modulation frequency, and a small impact parameter~\cite{gover2020free, ratzel2021controlling}. Second, decoherence mechanisms due to the quantum system's interaction with its environment may place even stricter constraints, necessitating electron beam drive rates faster than decoherence processes. Third, building experimental platforms that integrate well-controlled quantum systems with electron microscopy remains technically challenging~\cite{catanzaro2024experimental, jarovs2025sensing}. Recent experimental advances include observations of free-electron-induced decoherence mechanisms in hexagonal boron nitride color centers~\cite{taleb2024ultrafast}, as well as the integration of spin precession sensing with electron beams~\cite{jarovs2025sensing}. Nonetheless, no platform thus far has demonstrated resonant electron beam interactions with spin systems, nor have accurate estimations of free-electron--qubit interaction strength under realistic experimental conditions been made.

As a proof-of-concept step towards exploring free-electron--spin interactions, spin qubits -- specifically the negatively charged nitrogen-vacancy (\nv) center in diamond -- could be employed as quantum sensors to interrogate electron-beam interactions. A prerequisite to quantum control experiments would be to detect a signature of the modulated electron beam in the NV$^{-}$ resonant spectrum or in its characteristic decay times~\cite{degen2017quantum}. \nv~centers are one of the most widely used solid‑state spin‑qubit platforms, thanks to their room‑temperature operation, long coherence times, strong sensitivity to electric, magnetic, and strain fields~\cite{maze2008nanoscale, balasubramanian2009ultralong, doherty2013nitrogen, barry2020sensitivity, kolkowitz2015probing}, as well as straightforward readout and control schemes~\cite{jelezko2006single, rondin2014magnetometry}. Establishing such a fingerprint would be a critical first step toward control of solid‑state spin qubits with free electrons, while simultaneously providing a metrological benchmark for the free-electron--qubit coupling strength under realistic experimental conditions.

Here, we propose a general framework to realize quantum sensing of electron beams with solid-state spin qubits. We build an experimental platform to probe interactions between a modulated electron beam and the spin degree of freedom of optically active solid-state defects (here the \nv center in diamond). Our platform combines a GHz-bunched electron gun with a conventional confocal microscopy setup for optical readout of an ensemble of \nv centers. Within this platform we achieve four main milestones. First, we derive a first-principles master equation model of the magnetic free-electron--spin couplings. By including realistic qubit noise processes, we identify $T_{1}$ relaxometry as an experimentally viable route for observing this interaction and quantitatively describe the requirements for observing an experimental signature of free-electron--spin interactions. Second, by correlating cathodoluminescence with optically detected magnetic resonance (ODMR) measurements, we map beam-induced charge-state conversion, defining realistic operating windows for quantum sensing in our setup. Third, we perform systematic $T_{1}$ relaxometry in the presence of the beam, placing an upper bound on the free-electron--spin interaction. Finally, we translate these results into a roadmap that will guide future experiments to achieve quantum sensing of electron beams with single-spin qubits under experimentally realistic conditions. Collectively, these results establish the first quantitative benchmark for electron-beam quantum sensing with \nv centers and lay the groundwork for future quantum control and sensing experiments combining free electrons and spin qubits.

\begin{figure}
\centering
\vspace{-0.2cm}
  \includegraphics[scale=0.7]{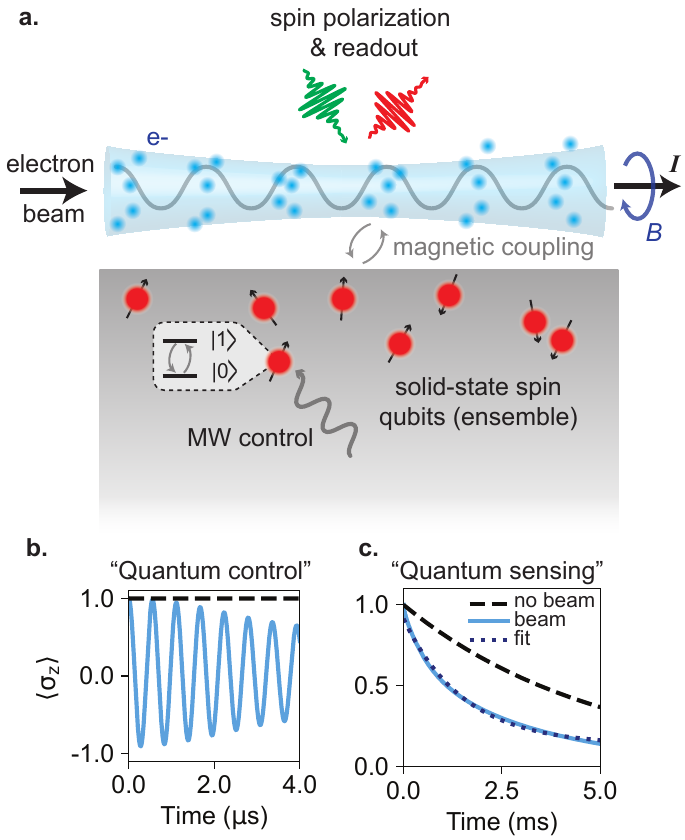}
    \caption{\small \textbf{Interactions between electron beams and solid-state spin qubits.} \textbf{a.} Conceptual schematic of the setup: a modulated beam of electrons impinges on a bulk crystal containing an ensemble of solid-state spins. The magnetic field generated by the beam couples to the spin. The spins can be initialized and read out through laser pulses, and controlled by an externally applied microwave drive field (MW). \textbf{b.} Predicted Rabi oscillation of the qubit (i.e. the spin manifold of the NV$^-$ center) from a resonantly modulated free-electron drive in the quantum control regime. We assume a perfectly bunched beam with average current $I=1$~mA, impact parameter $\rho_0 = 10~\si{\micro \meter}$, $T_1 = 5$~ms, $T_2=3.3$~ms, and an inhomogeneous decoherence rate $\gtwos$ of $1.9~$MHz. \textbf{c.} Predicted influence of the modulated electron beam on the effective energy relaxation time $T_1$ of a single NV$^-$ spin in the quantum sensing regime. We assumed a perfectly bunched beam with average current $I=16~\si{\micro \ampere}$, $T_2=100~\si{\micro \second}$, and the same inhomogeneous linewidth, $T_1$, and impact parameter. The estimated effective $T_1$ in the presence of the beam is reduced from $\sim5~\si{\milli\second}$ to $\sim1.5~\si{\milli\second}$.}
    \label{fig:concept}
    \vspace{-0.3cm}
\end{figure}

\begin{figure*}
\centering
\vspace{-0.2cm}
  \includegraphics[scale=0.54]{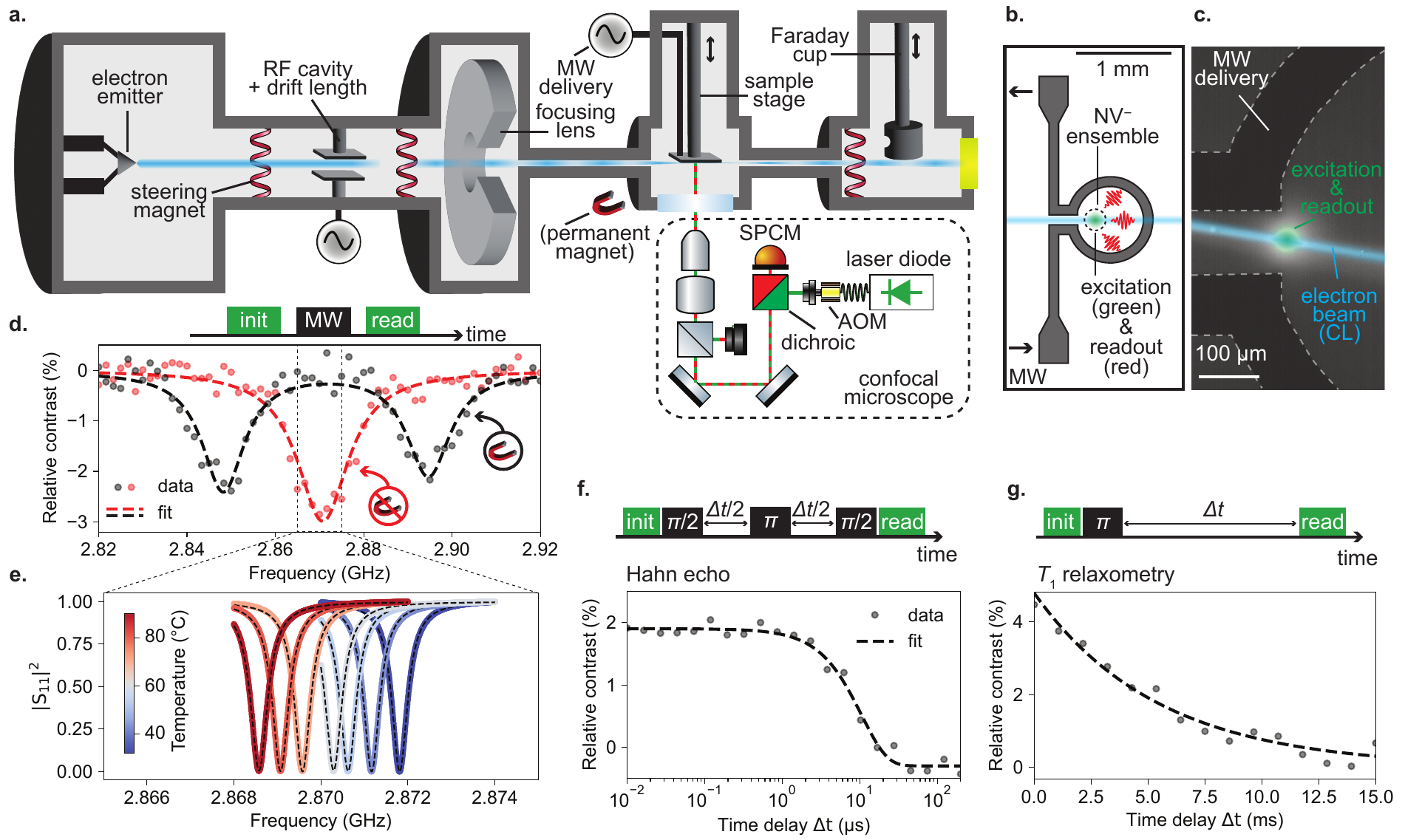}
    \caption{\small \textbf{Quantum sensing in a modulated electron beam line.} \textbf{a.} Schematic of the experimental setup, consisting of a bunched electron beam interacting with a spin system (as originally theoretically proposed in Ref.~\cite{ratzel2021controlling}). In addition, a confocal microscope with MW control is integrated to the beam line to initialize, control, and read out the spin state. \textbf{b.} Schematic of the NV$^-$ chip mounted on the sample stage. \textbf{c.} False-color optical micrograph of the chip under coincident electron beam and laser illumination. \textbf{d.} Optically detected magnetic resonance (ODMR) spectra in the absence (red) and presence (black) of an external permanent magnetic field, which splits the $\ket{0}\leftrightarrow\ket{-1}$ and $\ket{0}\leftrightarrow\ket{+1}$ transitions. The inhomogeneous linewidth observed here ($\approx14~\si{\mega \hertz}$) is broadened beyond $\gtwos \approx 12~\si{\mega \hertz}$, corresponding to an inhomogeneous linewidth of $\approx6~\si{\mega \hertz}$, due to a combination of MW power broadening and the finite pulse length used (see SI, Section S3). \textbf{e.} Thermally tunable bunching cavity resonance spectrum measured via its reflection coefficient $S_{11}$, approximately on resonance with the zero-field spin transition. \textbf{f.} Hahn echo sequence signal. From the fit we extract $T_2=10.6(9)\, \si{\micro \second} \approx\gtwo^{-1} $, where the value inbetween brackets denotes the standard error on the fit. \textbf{g.} Relaxometry pulse sequence in the absence of the beam. From the fit we extract $T_1=5.7(3)\,\si{\milli \second} = 1 / (2\gone)$.}
    \label{fig:setup}
    \vspace{-0.3cm}
\end{figure*}

\section{Results}

\section{Interactions between modulated electron beams and solid-state spins}

We derive a first-principle, fully quantum-mechanical description of the interaction between a modulated electron beam and a solid-state spin. We include realistic decoherence pathways for the spin, such as homogeneous and inhomogeneous transverse decoherence rates, and energy relaxation. In this section we summarize the main results (see Supplementary Information (SI), Section~S1 for more details).

We consider an electron beam (average current $I_0$) focused near (impact parameter $\rho_0$) the surface of a substrate containing (optically active) solid-state spins (Fig.~\ref{fig:concept}). The interaction can be described by considering the magnetic field generated by the mean electron current $I(t)$, and its (Poissonian) quantum fluctuations $\delta I(t)$, acting on the spin. Although we find that the latter introduces a direct spin decay pathway, its effects can typically be neglected for the macroscopic beam currents considered here (see SI Section S1). However, a much stronger, resonant interaction can be engineered by modulating the electron beam current: $I(t) = I_0 +  I_\mathrm{res}\cos{(\omega_I \, t)}$ to match the qubit transition frequency $\omega_0$ (corresponding to the $\ket{0}\rightarrow\ket{\pm1}$ transition, whose degeneracy may be lifted with a DC magnetic field, as shown in Fig.~\ref{fig:setup}). Then, under the rotating-wave approximation, the Hamiltonian is given by:
\begin{equation}\label{eq:Hamiltonian}
    H = \frac{\delta}{2} \sigma_z +  \frac{I_\text{res}\phi_0}{2e} \sigma_x \, ,
\end{equation}
with $\sigma_x, \sigma_z$ the Pauli matrices and $\delta = \omega_I - \omega_0$ a detuning parameter. On resonance ($\delta = 0$), the electron beam will drive the qubit with Rabi frequency $\rabi=I_\text{res}\phi_0/e$, given by the dimensionless parameter $\phi_0=\alpha\lambda_C/(2\pi\rho_0)$, where $\alpha$ is the fine structure constant and $\lambda_C$ the Compton wavelength, the electron charge $e$, and the beam impact parameter $\rho_0$ (related to the beam diameter). 

We incorporate homogeneous dephasing processes as well as energy relaxation of the spin by evaluating the master equation for the spin density matrix $\rho_s$:
\begin{align}\label{eq:master_equation}
    \frac{d\rho_s}{dt} = - \frac{i}{\hbar} \left[ H, \rho_s \right] + \mathcal{D}_\text{em}[\rho_s] + \mathcal{D}_\text{abs}[\rho_s] + \mathcal{D}_\text{dep}[\rho_s],
\end{align}
where the $\mathcal{D}$'s are dissipators associated with quantum jump operators $L_\text{em}=\sqrt{\gamma_1}\sigma_{-}$, $L_\text{abs}=\sqrt{\gamma_1}\sigma_{+}$, $L_\text{dep}=\sqrt{\gamma_2}\sigma_{z}$ and $\gone$ and $\gtwo$ longitudinal and transverse relaxation rates, respectively. Additionally, we include inhomogeneous dephasing by treating $\delta$ in Eq.~(\ref{eq:Hamiltonian}) as a Gaussian random variable drawn from a normal distribution with standard deviation $\sigma_{\omega_0} = \sqrt{2} \gtwos$, with $\gtwos$ the inhomogeneous dephasing rate. 
The derivation of the master equation is presented in the SI, Section~S1, and the reader is also referred to recent related work \cite{ratzel2021controlling,yuge2023super,taleb2024ultrafast}.

To gain insight into the spin dynamics under electron beam drive, we consider two experimentally relevant example parameter regimes, which we name \textit{quantum control} and \textit{quantum sensing} (Fig.~\ref{fig:concept}b,c). The \textit{quantum control} regime, defined by $\rabi \gg \gtwos,\gtwo,\gone$ enables full-contrast single-qubit quantum control using the electron beam, an important precursor for quantum information processing protocols (Fig.~\ref{fig:concept}b), and we discuss system requirements to achieve this regime in the Discussion section. 


In the remainder of this work, we investigate the \textit{quantum sensing} regime, which we define by: $ \gtwos \geq \rabi \gtrsim \gone$. This regime is particularly interesting for sensing applications such as the non-destructive detection and inspection of electron beams. In particular, we investigate the case: $\gtwos \gg \gtwo \gg \rabi \gtrsim \gone$, which can be considered the minimal requirement for detecting the free-electron--qubit interaction. To this end, we propose a sensing scheme that probes spin relaxation under resonant modulation of the electron beam ($\gamma_1^\text{beam}$), which we find yields an enhanced spin relaxation rate (compared to the intrinsic spin relaxation rate in the absence of the beam $\gamma_1$, with $T_1=1/(2\gone)$):
\begin{equation} \label{eq:T1_reduction}
    \gamma_1^\text{beam} \approx \gone +  \frac{\pi I_\text{res}^2\phi_0^2}{e^2} \,V(0,\sqrt{2}\gtwos, \gamma_2)\,,
\end{equation}
with $V(x,\sigma,\gamma)$ the Voigt profile function evaluated at $x$, with Gaussian standard deviation parameter $\sigma$ and Cauchy distribution parameter $\gamma$. This expression, a key result of the work, incorporates all relevant electron beam and qubit parameters, and enables direct assessment of the viability of observing an experimental signature in realistic settings. Specifically, our scheme shows that spin relaxometry is sensitive to the ratio $I_\text{res}/\rho_0$. In particular, for a typical \nv center ensemble, our relaxometry scheme relaxes the electron beam current requirements compared to resonant driving by two orders of magnitude, while still producing a clear signature on the effective $T_1$ decay time of the qubit (Fig.~\ref{fig:concept}d shows a reduction by a factor of $\approx 3.3$). For this reason, we propose spin relaxometry as the most promising near-term pathway for achieving the quantum sensing of electron beams. 


\section{Quantum sensing in an electron beam line}

Next, we describe our experimental efforts towards observing the free-electron--spin interaction, starting by introducing the platform shown in Fig.~\ref{fig:setup}a. We integrate a home-built confocal microscope into a bunched electron beam line. The beam line consists of a thermionic electron emitter accelerated to 10~keV, aligned through the chamber with steering magnets and focused down to a $\approx10~\si{\micro \meter}$ focal spot with a solenoid lens on a diamond bulk sample, which is common for such large current beam lines ($>\si{\micro \ampere}$). The beam is bunched by a GHz cavity followed by a $\approx 50$~cm drift length, realizing close to perfect bunching (see estimation of the bunching parameter in SI, Section~S2). The sample can be aligned with several translation stages in the vacuum chamber, and may be retracted to measure the average current on a Faraday cup further downstream. 

To polarize and read out the \nv~spin ensemble, we use a combination of optical pulses delivered by a 520~nm green laser diode modulated by an acousto-optic modulator or a laser diode control unit. Optical excitation is synchronized with MW pulses delivered by an $\Omega$-shaped wire deposited at the surface of the sample (see Methods and sample schematic in Fig.~\ref{fig:setup}b). Emission from the \nv~ensemble is collected with a single-photon counting module. The focal points of both the optics and the electron beam can be adjusted to coincide on the sample within the $\Omega$-shaped wire (see Fig.~\ref{fig:setup}c).  

We first benchmark the \nv~center ensemble with conventional pulse sequences used in spin control experiments~\cite{sewani2020coherent, rosenthal2023microwave}. We detect a photoluminescence dip (ODMR) when applying a field with the $\Omega$-shaped wire at the resonance frequency ($\omega_0 \approx 2.87~\si{\giga \hertz}$), which can be Zeeman-split by placing a permanent magnet in the vicinity of the sample chamber. Tuning the MW drive duration, we observe several cycles of Rabi oscillation, which demonstrates our ability to coherently drive the spin system (see SI, Section~S2). Additional pulse sequences to estimate $\gone = 88(5) ~\si{\hertz}, \gtwo=94(8)~\si{\kilo\hertz}$ and $\gtwos=12(2)~\si{\mega \hertz}$ are shown in Fig.~\ref{fig:setup}e,f and in the SI, Section~S2.

We also characterize the influence of the bunching cavity on the electron beam. The cavity is thermally tuned to be in resonance with the spin qubit (see Fig.~\ref{fig:setup}d), and we estimate the cavity generates an approximately perfectly bunched beam~\cite{ratzel2021controlling} (see SI, Section~S2). 
Cathodoluminescence imaging reveals that the cavity modulates the beam transversely at \(2\omega_I\), producing a double-peak profile, with optimal bunching achieved at approximately 15~dBm drive power without excessive transverse spreading (see SI, Section~S2).

\begin{figure*}
\centering
\vspace{-0.2cm}
  \includegraphics[scale=0.65]{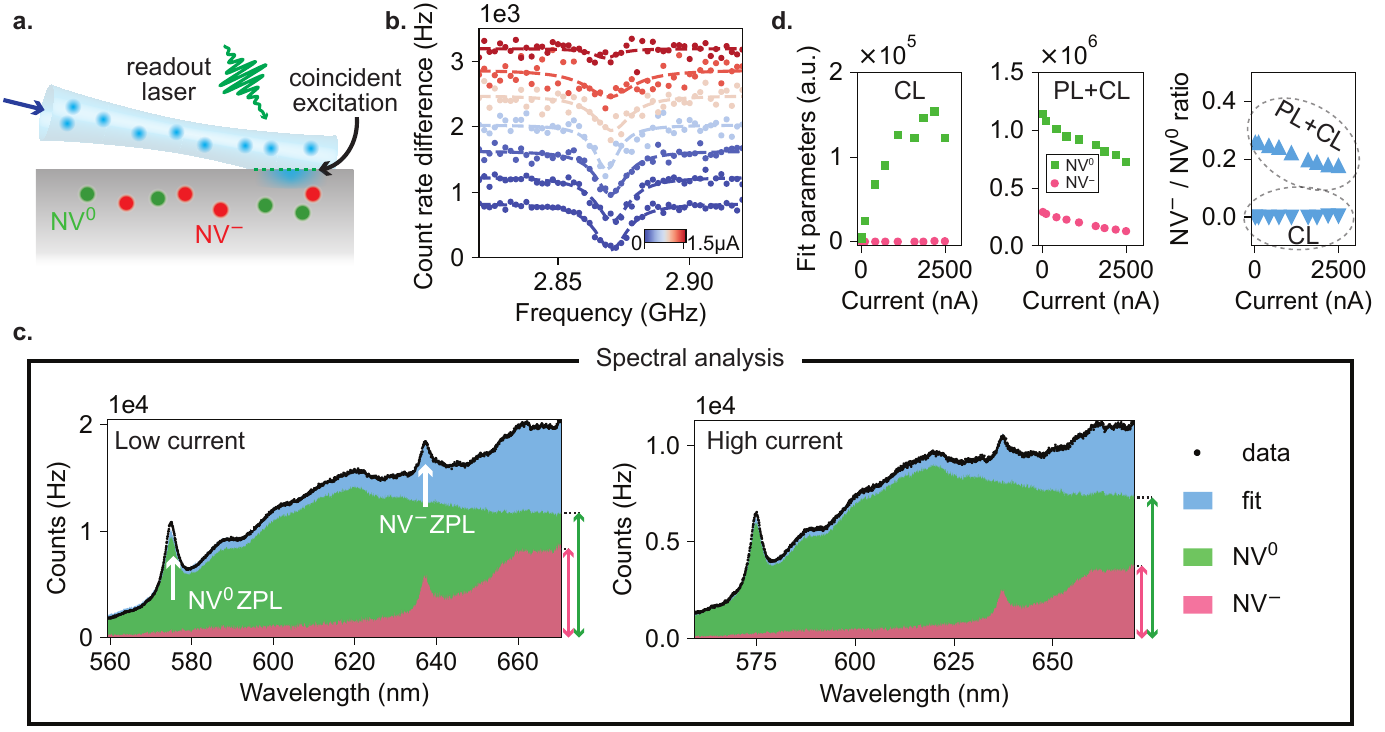}
    \caption{\small \textbf{Influence of charge conversion on spin ensemble readout.} \textbf{a.} Schematic of the beam exposure on a bulk sample with an equilibrium population of neutrally-charged (NV$^{0}$, green) and negatively-charged (NV$^{-}$, red) centers. The laser excitation for spin readout is co-located with the electron beam at the surface of the sample. \textbf{b.} ODMR spectrum as a function of electron beam current (absolute contrast, offset for clarity). \textbf{c.} Spectral analysis : a weighted sum of \nv and NV$^0$ contributions is fitted to the emission spectrum to estimate the ratio between the charge populations. Arrows on the right side of the plot show the relative weight of each charge state. \textbf{d.} Extracted weights from the spectral analysis in \textbf{(c)}, showing the relative weight of NV$^{0}$ and NV$^{-}$ as a function of beam current (for CL spectrum or coincident CL + PL).}
    \label{fig:cc}
    \vspace{-0.3cm}
\end{figure*}

\section{Charge stability of the \nv ensemble under electron beam irradiation}

The demanding requirements on the close proximity of the electron beam (see Eq.~(\ref{eq:T1_reduction})), inevitably lead to leaking of electrons into the diamond substrate. This may locally alter the charge environment, which we now experimentally study by co-located photoluminescence (PL) and cathodoluminescence (CL) measurements under electron beam exposure.

The schematic of the experiment is shown in Fig.~\ref{fig:cc}a. This allows us to measure the spin readout contrast as we turn up the beam current to currents where $T_1$ reduction should become observable ($>1~\si{\micro \ampere}$). We observe a significant decrease of the ODMR contrast (Fig.~\ref{fig:cc}b). Specifically, the absolute difference in counts with and without MW drive at the resonant frequency decreases by a factor of $\approx5$, as the current goes up to 1.5~$\si{\micro \ampere}$. 

Although this decrease in contrast is in part due to permanent sample damage (likely due to decreased collection efficiency as a result of surface graphitization), we will show here that charge conversion from the NV$^-$ to NV$^0$ also plays an important role. We measure CL (green laser off, electron beam on) and CL + PL (coincident green laser and electron beam) spectra (see raw spectra in SI, Section~S4) with increasing current. Cathodoluminescence spectra reveal that the electron beams preferentially excite the \nz~charge state~\cite{sola2019electron}, whose zero-phonon line is at $\approx 575$~nm. By performing a spectral analysis of the signal, we estimate the relative concentration of both charges states~\cite{schreyvogel2014tuned} (see Fig.~\ref{fig:cc}c, Methods, and SI, Section~S4). This spectral analysis reveals that PL + CL spectra contain a combination of both charge states. Interestingly, the ratio of \nv~to \nz~decreases as the current increases~\cite{sola2019electron} (see Fig.~\ref{fig:cc}d). Since \nz~states do not contribute to the ODMR signal, this charge conversion partially accounts for the reduction in ensemble readout contrast.

Importantly, this charge conversion analysis allows us to estimate a workable range for quantum sensing experiments in our experimental configuration, since the readout contrast decreases significantly at a few $\si{\micro \ampere}$, where the beam at grazing incidence ends up hitting the sample and delivering charge to the bulk. 

\section{$T_1$ relaxometry of electron beams}

\begin{figure}
\centering
\vspace{-0.2cm}
  \includegraphics[scale=0.6]{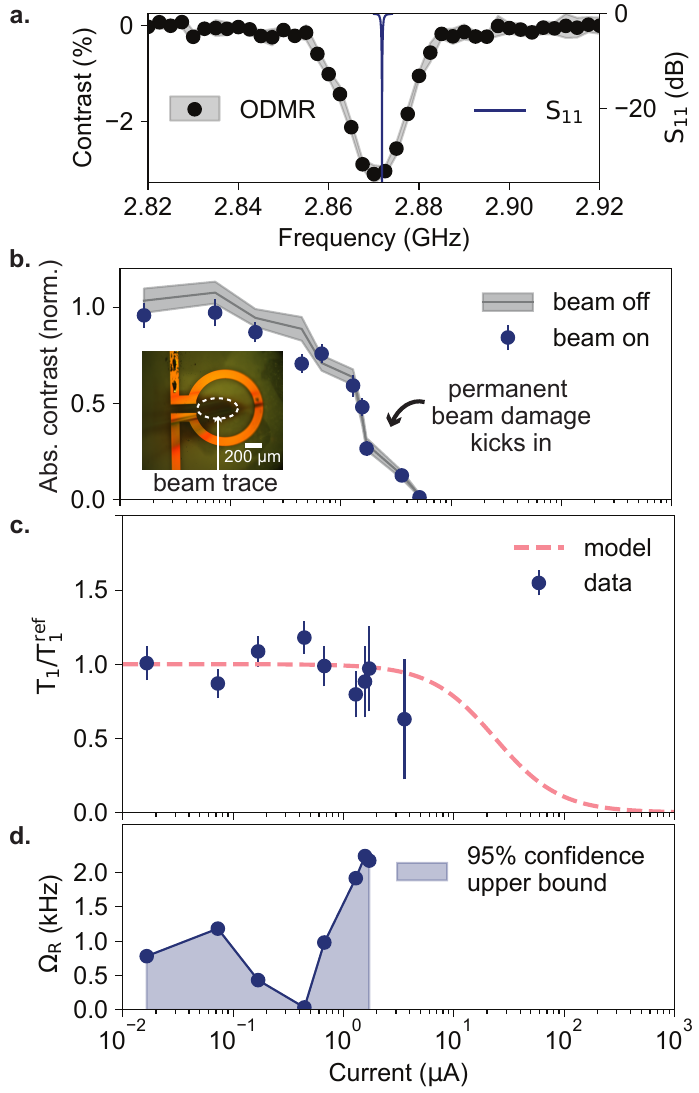}
    \caption{\small \textbf{$T_1$ relaxometry of modulated electron beams.} \textbf{a.} ODMR and cavity resonance spectra for resonant free-electron--qubit conditions (as in Figure \ref{fig:setup}d). \textbf{b.} Absolute contrast as a function of electron beam current, normalized to a reference measurement in the absence of electron beam taken at the beginning of the sweep. The gray area corresponds to the average absolute contrast ($\pm \sigma$) of a reference measurement (electron beam off) taken immediately after each data point. Inset: optical micrograph of the sample after prolonged beam exposure, showing permanent beam damage (also see SI, Section~S4). \textbf{c.} $T_1$ relaxometry as a function of beam current. Each point corresponds to the ratio of the measured $T_1$ under electron beam excitation to a reference $T_1$ measurement (in the absence of electron beam, taken at the beginning of the sweep). Error bars indicate the standard error on the fit of a single exponential function to the data (here and in (\textbf{b}). \textbf{d.} Upper bound on the free-electron--qubit coupling strength. Points that yielded negative bound values (due to the large error bars) are not shown.}
    \label{fig:t1s}
    \vspace{-0.3cm}
\end{figure}

Next, we perform $T_1$ relaxometry of the \nv ensemble under the influence of the electron beam. We operate in the low-field regime (degenerate spin transitions) and tune the cavity resonance to match the ODMR peak (see Fig.~\ref{fig:t1s}a-b). We compare $T_1$ measurements with a single reference measurement (beam off, $T_1^\text{ref}$) and estimate the ratio $T_1/T_1^\text{ref}$ (see Methods and raw data in the SI, Section~S3). We also interleave reference $T_1$ measurements with the beam turned off to monitor permanent beam damage to the sample.

We gradually increase the average current up to $\approx 3.5~\si{\micro \ampere}$ when permanent beam damage, likely due to graphitization of the sample surface (see Fig.~\ref{fig:t1s}b), combined with the effects of charge conversion, results in a decreased signal-to-noise ratio, and hence a greater uncertainty on the measured $T_1$. Up to these currents we observe no significant decrease of the effective $T_1$ time, which is consistent with our theoretical model (see Fig.~\ref{fig:t1s}c). To reach the regime of significant $T_1$ reduction ($<0.5$), the current should be increased to $\gtrsim 10~\si{\micro \ampere}$, which is currently unattainable due to the significant decrease in readout contrast. We also conduct experiments in the non-resonant condition by Zeeman-splitting the spin transition with an external field while fixing the cavity resonance and observe no significant $T_1$ decay, as expected (see SI, Section~S3).

Nevertheless, our $T_1$ relaxometry protocol can still provide a bound to the strength of the free-electron-qubit interaction according to our theoretical model. Specifically, we calculate the 95\% confidence interval lower bound of the $T_1/T_1^\text{ref}$ ratio, which can be interpreted as an upper bound on $\rabi = I_\text{res}\phi_0/e$ (see Fig.~\ref{fig:t1s}d and SI, Section~S3). The values we obtain are consistent with the currents measured at the Faraday cup. We also note that other factors may reduce the effective current contributing to the interaction: (1) the electron beam may not be perfectly delivered onto the sample due to charging effects and focusing artifacts in the solenoid lens; (2) although the beam is theoretically perfectly bunched, the parasitic transverse deflection reduces the effective current (compared to the unmodulated condition) by $\approx 35\%$. Future enhancements in the spin coherence and electron beam delivery, which we discuss in the next section, may enable the regime of strong $T_1$ reduction and yield tighter bounds on the free-electron--qubit coupling strength.

\begin{figure}
\centering
\vspace{-0.2cm}
  \includegraphics[scale=0.6]{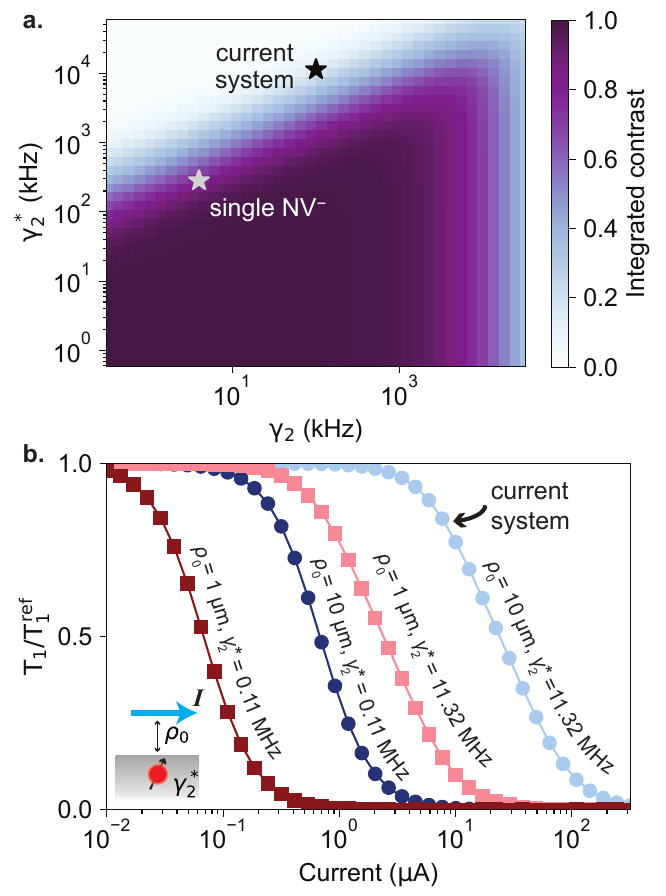}
    \caption{\small \textbf{A roadmap for quantum sensing of electron beams with spin qubits.} \textbf{a.} Numerical simulation of the integrated relaxometry contrast between an electron beam-driven system and non-driven system as a function of inhomogeneous and homogeneous transverse relaxation rates $\gtwos$ and $\gtwo$. Stars indicate parameter values of our current system and an ideal single \nv~system. We assume a drive current of $10~\si{\micro \ampere}$ and impact parameter $\rho_0=10~\si{\micro \meter}$. Single \nv parameters are taken from Ref.~\cite{van2012decoherence}. \textbf{b.} Predicted $T_1$ reduction as a function of electron beam current for different electron beam configurations and inhomogeneous spin qubit decay rates. The rightmost line corresponds to our current system, while the other lines correspond to reductions in the inhomogeneous transverse relaxation rate and reduction of the impact parameter. We assume $T_1=5~$ms and $T_2=10~\mu$s.}
    \label{fig:vision}
    \vspace{-0.3cm}
\end{figure}

\section{Discussion} 

Our framework singles out two main avenues for observing an experimental signature of the interaction between electron beams and spin systems: (1) optimizing spin performance and (2) improving electron beam delivery. 

Readily available color center material platforms with greater coherence may already achieve such enhancements: for instance, in Fig.~\ref{fig:vision}a, we show the expected integrated contrast (comparing the $T_1$ decay with and without electron beam, see Methods) as a function of homogeneous and inhomogeneous transverse decay rates $\gtwos$ and $\gtwo$. Our experiment is most sensitive to the inhomogeneous decay rate, such that an improvement in $\approx 1$ order of magnitude in $\gtwos$ would already result in close-to-unity contrast at the currents realized in our experiment. Such lower inhomogeneous broadening has previously been recorded in bulk \nv~ensembles at lower concentrations~\cite{bauch2020decoherence}. Single spin qubits (e.g., at cryogenic temperatures) offer even greater performance and may be the platform of choice for future quantum sensing experiments with electron beams~\cite{abobeih2018one}.

Another avenue is to improve the electron beam delivery onto the sample, where the figure of merit is given by $I_\text{res}/\rho_0$ ($\lesssim0.1~\si{\ampere \per \meter}$ in our system). We expect that higher-brightness field-emission electron guns may achieve $I_\text{res}/\rho_0\approx \SI{1}{\ampere \per \meter}$ (see Fig.~\ref{fig:vision}b, pink and brown data and Section~S5 in the SI). Field-emission guns can realize these values at lower currents and impact parameters, where lower currents and improved collection optics can mitigate beam damage on the sample. It is worth noting that strategies for grazing interactions between electron beams and extended nanophotonic structures have been developed over the past decade~\cite{dahan2020resonant, kfir2020controlling}, and may be used to mitigate beam-induced charge conversion. Such grazing interaction schemes were also proposed for mediating interactions between free electrons and matter qubits~\cite{karnieli2023quantum, karnieli2024universal, abad2024electron}. Such methods should already enable the propagation of swift electron beams in close vicinity to a point-like spin system. Alternatively, recent multi-color pumping schemes~\cite{mahdia2025high} may also be used to mitigate charge conversion induced by electron beam.

Such enhancements in electron beam and color center coherence may enable us to access the regime of \textit{quantum control} (see Fig.~\ref{fig:concept}b). Whether such coherent control is possible in practice is set by the magnitude of $\rabi$ compared to the noise sources in the system. 
In state-of-the-art field-emission guns (e.g., with brightness $>10^9~\si{\ampere \per \square \centi\meter \per \steradian})$, one can achieve $I_\text{res}/\rho_0\gtrsim10~\si{\ampere \per \meter}$ at higher energies (80-100~keV) which, in conjunction with a perfectly bunched electron beam, could yield Rabi frequencies on the order of $>100~\si{\kilo \hertz}$, comparable to inhomogeneous qubit dephasing rates ($\gtwos\sim 0.1 - 1 \si{\mega \hertz}$)~\cite{van2012decoherence, abobeih2018one}, making complete control potentially attainable. 

The hybrid platform demonstrated here -- integrating spin-qubit spectroscopy within a bunched electron beam line -- opens a new diagnostic and control modality for the electron microscopy community~\cite{jarovs2025sensing}. Ensembles or single spin qubits may provide \textit{in situ} beam diagnostics, leveraging advanced magnetometry protocols~\cite{herb2025quantum}. Looking further ahead, the same architecture lays the groundwork for quantum control of solid-state spins by free electrons. A useful historical precedent is the rapid adoption of ultrafast laser techniques for cathodoluminescence spectroscopy~\cite{polman2019electron} and photon-induced near-field electron microscopy~\cite{Barwick2009}, which are now routine in state-of-the-art scanning and transmission electron microscopes for mapping nanophotonic modes and ultrafast dynamics. 

We believe the combination of spin-based quantum sensing and electron microscopy may follow a similar trajectory: once available as a turnkey module, electron-beam-qubit-photonic interfaces could augment electron microscopy’s structural and optical repertoire with quantum-mechanical probes and actuators. Our results provide an initial step toward that possibility, in which quantum sensing and control may one day integrate naturally into electron microscopes and beam lines~\cite{ruimy2025free}.

\section{Methods}

\textbf{Theoretical model.}
We model the spin system as an effective two‑level system ${\ket{0},\ket{1}}$ subjected to evanescent, time‑varying magnetic fields generated by the modulated electron beam. This yields the Lindblad master equation from the main text. The coherent drive $\rabi\propto I_\text{res}$ is set by the magnitude of the resonantly modulated current, and the relaxation rates are determined based on properties of the spin system, measured with standard quantum sensing pulse sequences. For the simulation results presented in Figs.~\ref{fig:concept} and \ref{fig:vision}, the master equation (Eq. \ref{eq:master_equation}) is integrated numerically in \textsc{QuTiP}~\cite{johansson2012qutip}. From the resulting trajectory we obtain $\langle\sigma_z(t)\rangle$ (used to extract $T_1$). More details about the theoretical and numerical methods can be found in the SI, Section~S1.

\textbf{Sample fabrication.} The sample used for microwave spin control is a diamond substrate with photolithographically defined electrical contacts. The bare diamond substrate is a CVD-grown single crystalline diamond doped with 300~ppb NV centers (DNVB1, Thorlabs). Prior to photolithography, the sample surface is acid treated to remove unwanted residue. An initial acetone and isopropyl alcohol (IPA) sonication is done to remove macroscopic debris. Afterwards, the sample is cleaned in a 3:1 sulfuric acid to hydrogen peroxide Piranha solution, and then subsequently in 51\% hydrofluoric acid solution. Following the acid clean, the sample is once more sonicated in acetone and IPA.

The liftoff layer is done with a photoresist bilayer to improve the edge definition of the final electrodes. The first layer is 220~nm PMGI SF5, and the entire sample is baked at 150$^\circ$C for 1~minute. The second layer is 2~$\mu$m S1812, and the entire sample is baked at 115$^\circ$C for 1 minute. Patterning is done via Heidelberg MLA 150, exposing with a dosage of 625~mJ/cm$^2$ with a 375~nm laser. After exposure, development is done via MF319, a TMAH-based developer.
The metal for the electrical contacts is deposited via electron beam evaporation using a Kurt J. Lesker LAB18 system. The full stack consists of a 10~nm Ti adhesion layer followed by a 200~nm Au layer. Liftoff is done in N-Methyl-2-pyrrolidone (NMP)-based solvent Kayaku Advanced Materials Remover PG solution heated at 80$^\circ$C. After liftoff, a final clean is done via acetone/IPA solvent spray. The sample is then mounted onto a printed circuit board (PCB) using vacuum-compatible silver paste and the chip contacts are connected to the PCB using several wirebonds with the Westbond wire bondEIR.

\textbf{Experimental setup.}
A $\text{LaB}_6$ thermionic electron emitter with a 15~\si{\micro\meter} flat optimized for producing moderate beam current at micron spot sizes (Kimball Physics EGH-8103 delivers a beam that is focused by a solenoid lens down to $\approx10~\si{\micro \meter}$ spot at the sample. We operate the gun at 10~keV acceleration voltage for optimal bunching. A retractable Faraday cup provides an \textit{in situ} calibration of the average current $I_0$ ($0-10~\si{\micro \ampere}$). Optical excitation of the diamond sample is performed in a confocal geometry: 1-5~mW of 520~nm light is focused onto the chip through a 0.42‑NA 20$\times$ objective; the resulting photoluminescence is collected by the same objective, spectrally filtered and detected by a fiber-coupled single-photon counting module. Microwave control is generated by a signal generator with internal IQ modulation (SRS~SG394), generating 0-9~dBm of power, amplified by $\approx 40~$dB and delivered to the on‑chip $\Omega$‑shaped wire. The beam is modulated longitudinally by a resonant radio-frequency (RF) klystron-like cavity. A small transverse modulation is also observed. Bunching parameter estimation can be found in the SI, Section~S2-3.

\textbf{Electron-beam alignment.} We use a combination of four steering magnets placed along the electron beam line to propagate the beam through the beam line (full length is $\approx 2$~m from gun to final scintillator screen). When performing quantum sensing experiments of the electron beam, we tightly focus the electron beam on the sample surface, as shown in Fig.~\ref{fig:setup}c. At high currents, sample charging may result in small beam drifts which we manually correct by keeping the beam aligned with the steering magnets on the green excitation laser. When switching between resonant and non-resonant excitation conditions, we need to bring a permanent magnet close to the sample stage, which affects the electron beam alignment. We then realign the beam on the sample in these new conditions.  

\textbf{$T_1$ relaxometry experiments.}
For each beam current setting we apply a $T_1$ relaxometry pulse sequence. The resulting fluorescence is fitted to $A\exp(-t/T_1)$ to extract $T_1$. Permanent beam damage results in low signal-to-noise ratio for the high current data points, therefore we did not include such experimental data (when the mean uncertainty is greater than the mean contrast). Statistical methods to evaluate the confidence interval and bound the current and interaction strength within some confidence interval are described in the SI, Section~S3.

\textbf{Charge‑conversion experiments.}
For each grid-voltage setting we measure (i) a pulsed ODMR spectrum and (ii) broadband emission spectra. For each current in (ii), we measure the CL only (green laser off) and the PL + CL spectra (overlapping green laser and electron beam on the sample). The CL and PL+CL spectra are decomposed into \nv~and NV$^0$ components by least‑squares fitting to a linear superposition of reference spectra. From this fit, we extract the relative contribution of \nv~and \nz~to the spectra. Additional details are provided in the SI, Section~S4.

\section{Authors contributions}
JMG, DC, CRC, EIR, and JV initially conceived the project. JMG, DC, and CRC built the electron beam line, with contributions from KJL and DSB. 
CRC, JMG, and EIR built the quantum sensing experiment, with contributions from GS, SB, and GLvdS. JMG and DC fabricated the sample. CRC and JMG performed the experiments. AK, GLvdS, CRC, and JMG developed the theoretical models and numerical analysis.
CRC, GLvdS, and AK analyzed the experimental data. 
CRC, RLB, IK, RJE, SF, OS, and JV supervised the research. CRC wrote the paper with inputs from all authors.

\section{Competing interests}
The authors declare no competing interest.

\section{Data and code availability statement}
The data and codes that support the plots within this paper and other findings of this study are available from the corresponding authors upon reasonable request. Correspondence and requests for materials should be addressed to chrc@stanford.edu.

\section{Acknowledgments}
The authors would like to thank Regina Lee, Ofer Kfir, and Joonhee Choi for stimulating discussions.

The work from the Vuckovic lab was supported
in part by a Vannevar Bush Faculty Fellowship from the US Department of Defense (DC, GS, SB) and the Gordon and Betty Moore
foundation (JMG, DC, CRC).
JMG acknowledges support from the Hertz Fellowship. CRC is supported by a Stanford Science Fellowship. GS acknowledges support from the Stanford Bloch Postdoctoral Fellowship. GvdS's salary was supported by the Department of Energy grant DE-SC0025295. AK is supported by the VATAT-Quantum fellowship by the Israel Council for Higher Education; the Urbanek-Chodorow postdoctoral fellowship by the Department of Applied Physics at Stanford University; the Zuckerman STEM leadership postdoctoral program; and the Viterbi fellowship by the Technion.

\bibliographystyle{ieeetr}
\bibliography{bibliography}

\end{document}